# Various arsenic network structures in 112-type Ca$_{1-x}$La$_x$Fe$_{1-y}$Pd$_y$As$_2$ revealed by synchrotron x-ray diffraction experiments

Shinya Tamura, Naoyuki Katayama[*], Yuto Yamada, Yuki Sugiyama, Kento Sugawara and Hiroshi Sawa

Department of Applied Physics, Nagoya University, Nagoya 464-8603, Japan

**ABSTRACT:** Two novel 112-type palladium doped iron arsenides were synthesized and identified using comprehensive studies involving synchrotron x-ray diffraction and x-ray absorption near edge structure (XANES) experiments. Whereas *in*-plane arsenic zigzag chains were found in 112-type superconducting iron arsenide, Ca$_{1-x}$La$_x$FeAs$_2$ with maximum $T_c$ = 34 K, deformed arsenic network structures appeared in 112-type materials such as longitudinal arsenics zigzag chains in CaFe$_{1-y}$Pd$_y$As$_2$ (y ~ 0.51) and arsenic square sheets constructed via hypervalent bonding in Ca$_{1-x}$La$_x$Fe$_{1-y}$Pd$_y$As$_2$ (x ~ 0.31, y ~ 0.30). As K-edge XANES spectra clarified the similar oxidization states around FeAs4 tetrahedrons, expecting us the possible parents for high $T_c$ 112-type iron arsenide superconductors.

## ■ INTRODUCTION

*P-block* elements construct various network structures that satisfy the Zintl-Klemm electron counting rule. Light *p*-block elements easily exhibit strong *s-p* hybridization to form complex network structures such as honeycomb lattices in MgB$_2$[1] and CaC$_6$[2] and diamond lattices in elemental Si. By contrast, heavy *p*-block elements such as Sb and Te show weak *s-p* hybridization, resulting in lattice networks oriented by strong *p* character such as zigzag chains in CaSb$_2$[3] and EuSb$_2$[4]. Furthermore, some compounds with heavy *p*-block elements often exhibit hypervalent states[5] that do not obey the Zintl-Klemm electron counting rules, resulting in the formation square sheets and ladders.

The bonding chemistry exhibited by *p*-block elements can potentially enable the creation of novel parents for high $T_c$ iron-based superconductors. The iron based superconductors consist of the alternate stacking of FeAs layers and the spacer layers. Recently, arsenic network structures have been introduced as spacer layers in, for example, Ca$_{10}$($M_4$As$_8$)(Fe$_{2-x}$$M_x$As$_2$)$_5$ (*M* = Pt, Ir, and Pd) with maximum $T_c$ = 38 K, which contains Fe$_2$As$_2$ and $M_4$As$_8$ spacer layers in which neighboring *M*As$_4$ squares are connected through either corner sharing or dimerization between adjacent divalent arsenics[6-12]. Another example is the 112-type iron arsenide, Ca$_{1-x}$La$_x$FeAs$_2$ with maximum $T_c$ = 34 K, in which adjacent monovalent arsenic atoms form zigzag chain layers[13-16]. Considering that As is located between light elements such as Si and heavy elements such as Sb and Te on the periodic table, the emergence of the various arsenic network structures can be expected in iron arsenides depending on the degree of *s-p* hybridization, leading to novel parents for high $T_c$ superconductivity.



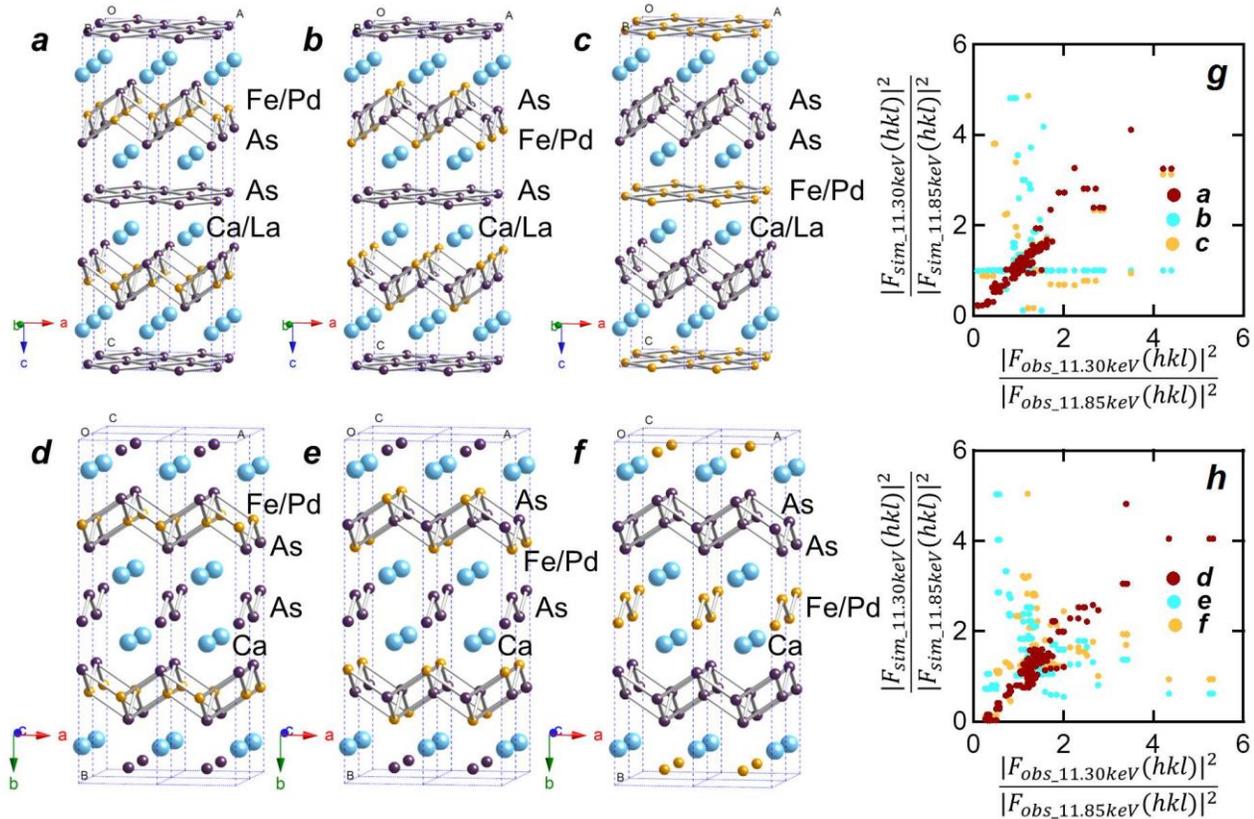

Figure 1. The possible crystal structures for tetragonal (*a-c*) and orthorhombic (*d-f*) samples obtained from conventional structural analysis using single crystalline x-ray diffraction data. The obtained reliability factors are R1 = 7.17, 7.34, and 7.42 for the respective tetragonal samples. For the orthorhombic samples, R1 = 3.67, 5.14, and 4.74 were attained, respectively. The relative ratios of the structural factors with the incident x-ray energies of 11.30 and 11.85 keV compared between the simulated and experimental results in *g* for tetragonal and *h* for orthorhombic samples clearly indicate a positive relation by assuming *a* for tetragonal and *d* for orthorhombic structures, respectively. See text for more details.

To address this issue, we focus our attention on 112-type families. Although 112-type iron arsenides have never been identified except for the superconducting $Ca_{1-x}La_xFeAs_2$ and its derivatives with zigzag chain layers[13-16], there exist in other 112-type families many isostructural materials with a variety of network structures composed of *p*-block elements, *e.g.*, *in*-plane zigzag chains in $SmAuAs_2$[17], cis-trans chains in $LaAgAs_2$[17], antimony square sheets constructed by hypervalent bonding in $BaZnSb_2$[18], and the longitudinal zigzag structure that is believed to connect neighboring layers in $CeAgSi_2$[19]. Thus, we can expect further possibilities to obtain novel 112-type iron based families from the proper tuning of arsenic network structures. This will lead to opportunities for the comparative study of the relationship between electric states and their underlying lattice geometries.

In this paper, we report two unprecedented 112-type Pd doped iron arsenides with various arsenic network structures as spacer layers. The substitution of Pd for Fe made it difficult to identify elemental species using conventional x-ray diffraction experimental methods because the atomic scattering factors of As and Fe/Pd ions are, by coincidence, nearly equivalent. However, by taking advantage of the energy dependent terms of the atomic form factors, or so-called anomalous dispersion terms, were able to successfully distinguish the scattering factors Additionally, we determined through x-ray absorption near edge structure (XANES) experiments that the 112-type iron arsenides have similar electronic states. We also discuss the dominant factors causing deformation of arsenic network structures in 112-type iron-based arsenides.

■ **EXPERIMENTAL SECTION**

**Synthesis.** The polycrystalline samples were obtained by heating a mixture of Ca, La, Pd, FeAs, and As powders with various nominal compositions. We will discuss the details of the respective compositions later. The mixture was placed in an aluminum crucible and sealed in an evacuated quartz tube. The preparation was then carried out in a glove box filled with nitrogen gas,



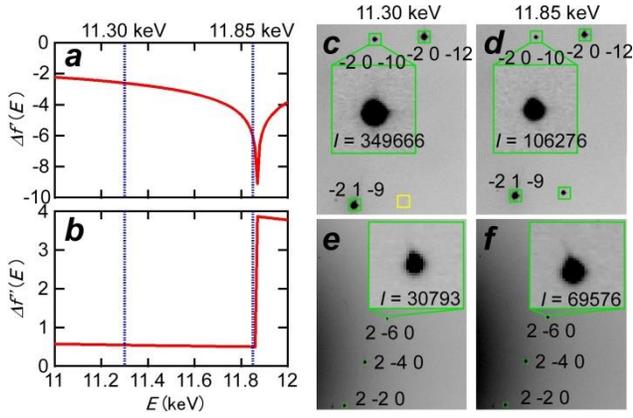

Figure 2. *a* and *b* show the energy dependences of the anomalous dispersion term of the atomic form factor for As near the K absorption edge. *c* and *d* show the energy dependences of the 20–10 Bragg peak intensity for the tetragonal samples, while *e* and *f* indicate the energy dependences of the 2–60 Bragg peak intensity for the orthorhombic samples. *c* and *e* are measured at 11.30 keV, while *d* and *f* are measured at 11.85 keV.

with the ampules heated at 700°C for 3 h and then slowly heated further up to 1,100°C, cooled to 1,050°C at a rate of 1.25°C/h, and finally furnace cooled.

**Synchrotron x-ray diffraction.** Single-crystals with a typical dimension of $60 \times 50 \times 30$ μm$^3$ were picked from the obtained polycrystalline samples and used for synchrotron x-ray diffraction experiments at the BL8A beamline at PF, the KEK facility in Tsukuba, Japan. The obtained lattice parameters and refined conditions are summarized in Table I.

**XANES experiment.** XANES spectra were then obtained using the same $60 \times 50 \times 30$ μm$^3$ single crystals that were used for the single crystal x-ray diffraction experiments. The XANES experiments were conducted at the BL11S2 of Aichi Synchrotron Radiation Center of the Aichi Science & Technology Foundation in Aichi, Japan. The obtained data were analyzed using the Athena software package.

■ **RESULTS AND DISCUSSIONS**

**Crystal structural analysis.** A synchrotron x-ray with an incident energy of $E = 18.0$ keV was used to conduct structural analysis. Although diffuse scatterings appeared, all of the peaks were distinct and could be successfully indexed as sharp Bragg peaks, indicating the presence of high quality single crystals without impurities. We will discuss the origin of diffuse scatterings later. Although superconducting $Ca_{1-x}La_xFeAs_2$ crystallizes in a monoclinic structure with typical lattice parameters of $a = 3.94710(10)$ Å, $b = 3.87240(10)$ Å, $c = 10.3210(7)$ Å, and $\beta = 91.415(6)°$ [13], we obtained tetragonal and orthorhombic arsenides in these studies, which we will hereafter refer to as "monoclinic samples," "tetragonal samples," and "orthorhombic samples," respectively. After careful investigation of extinction rules and Laue symmetry, we determined the space group to be *I*4/*mmm* and *Cmcm* for the tetragonal and orthorhombic samples, respectively. As a result of the ionic size effect, the *in*-plane lattice parameters were increased by substituting Pd for Fe.

Following the conventional procedures for structural refinement, we successfully determined the atomic positions for both samples. However, as shown in Figure 1, there existed three possible candidate structures, each with a low reliability factor, *R1*, for each sample owing to the close similarity between the atomic form factors of As and Fe/Pd ions, which means that the relation $f_{As} \cong (1-x)f_{Fe} + xf_{Pd}$ is satisfied. Although the

Table I. Summary of collected and refined data for tetragonal and orthorhombic samples at 120 K. The data were collected using a synchrotron x-ray with an incident energy of $E = 18.0$ keV.

| A. | Tetragonal Sample |
|---|---|
| Chemical Formula | $Ca_{1-x}La_xFe_{1-y}Pd_yAs_2$ |
| Space group | *I*4/*mmm* |
| cell dimensions | |
| $a$ (Å) | 3.99280(10) |
| $c$ (Å) | 20.4154(7) |
| $V$ (Å$^3$) | 325.472(16) |
| observed reflections | 2772 |
| resolution (Å) | $d > 0.40$ |
| independent reflections | 721 |
| Completeness | 0.846244 |
| Redundancy | 3.844660 |
| $R$ merge | 5.934600 |
| $R$ merge (2σ cut) | 5.911445 |
| B. | Orthorhombic Sample |
| Chemical Formula | $CaFe_{1-y}Pd_yAs_2$ |
| Space group | *Cmcm* |
| cell dimensions | |
| $a$ (Å) | 4.07440(10) |
| $b$ (Å) | 17.6291(4) |
| $c$ (Å) | 4.1520(7) |
| $V$ (Å$^3$) | 298.23(5) |
| observed reflections | 3057 |
| resolution (Å) | $d > 0.40$ |
| independent reflections | 1231 |
| completeness | 0.861442 |
| redundancy | 2.483347 |
| $R$ merge | 2.654790 |
| $R$ merge (2σ cut) | 2.587448 |



models depicted in Figures 1*c* and 1*f* could be safely excluded based on chemical considerations, further experiments were required to determine which structures were assumed by the respective samples. To make the distinctions, we employed the energy dependences of the Bragg peak intensity around the absorption edge. In general, the atomic form factor can be expressed as

$$f = f^0 + \Delta f' + i\Delta f'' \qquad (1)$$

where $f^0$ is the *Q* dependent term and $\Delta f' + i\Delta f''$ is the anomalous dispersion term dependent on *E*. Because $\Delta f'$ shows divergence and $\Delta f''$ jumps discontinuously at the absorption edge, as shown in Figures 2*a* and 2*b*, we can observe the strong *E* dependences on Bragg peak intensity around the As K absorption edge. This *E* dependence allowed us to distinguish As sites from Fe/Pd sites, enabling the identification of all atomic sites.

Figures 2*c*-2*f* show the single crystal x-ray diffraction experimental patterns produced at incident x-ray energies of *E* = 11.30 and 11.85 keV for tetragonal (2*c* and 2*d*) and orthorhombic samples (2*e* and 2*f*). Note that incident x-ray energies higher than the As K absorption edge energy of 11.8667 keV were not used in this experiments to avoid statistical degeneration in the data as a result of the strong absorption of incident x-rays and subsequent fluorescence at such high energies. The relative ratios of the structural form factors to incident x-ray energy can be estimated using the following equation:

$$\frac{I_{11.30keV}(hkl)}{I_{11.85keV}(hkl)} = \frac{A_{11.30keV}(hkl)|F_{obs\_11.30keV}(hkl)|^2}{A_{11.85keV}(hkl)|F_{obs\_11.85keV}(hkl)|^2} \quad (2)$$
$$\cong \frac{|F_{obs\_11.30keV}(hkl)|^2}{|F_{obs\_11.85keV}(hkl)|^2}$$

where *I* and *F* indicate the Bragg peak intensity and the structural form factor at a specific incident energy, respectively, and *A* is a function of the absorption coefficient and the Lorentz and polarization factors. The approximation $A_{11.30keV} \cong A_{11.85keV}$ can be applied when performing diffraction experiments with different incident energies under otherwise similar conditions. The obtained relative ratios were then compared with simulated relative ratios calculated using

$$\frac{|F_{sim\_11.30keV}(hkl)|^2}{|F_{sim\_11.85keV}(hkl)|^2} = \frac{\left|\sum_j f_{j_{11.30keV}} e^{-2\pi i(hx_j+ky_j+lz_j)}\right|^2}{\left|\sum_j f_{j_{11.85keV}} e^{-2\pi i(hx_j+ky_j+lz_j)}\right|^2} \quad (3)$$

where *j* indicates the elemental specie. Both experimental and simulated ratios were collected for all indices available in these experiments. As summarized in Figures 1*g* and 1*h*, a clear positive relation was obtained only when the structures depicted in Figures 2*a* and 2*d* were assumed, indicating that these structures are the ones occurring in the analyzed samples. The resulting refinement details, including atomic coordinates, are summarized in Table II.

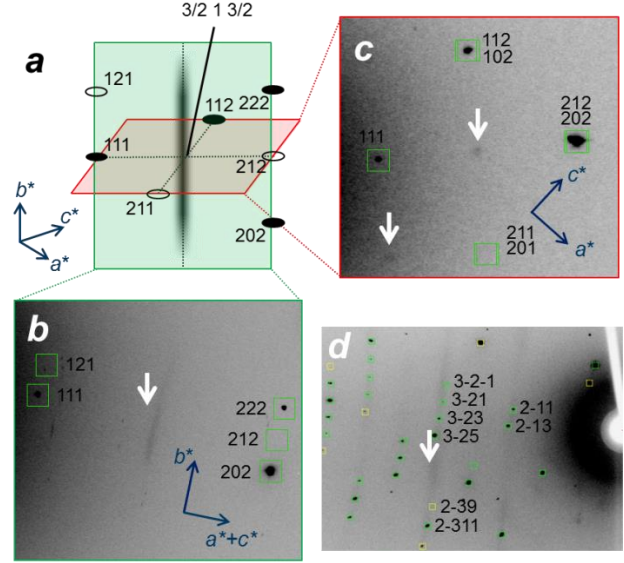

Figure 3. *a* shows a schematic of the diffuse streak around the 3/2 1 3/2 position in the orthorhombic samples. Thus was constructed via several diffraction experiments, as shown in *b* and *c*. The closed and open circles indicate the Bragg and extinction peaks, respectively. *d* indicates the diffuse scattering occurring in the tetragonal samples.

**Diffuse scattering.** The single crystal x-ray diffraction experiments summarized above clarified that temperature-independent diffuse scattering peaks appeared in both the tetragonal and orthorhombic samples, indicating the inherent short-range ordering emerges shown in Figure 3. By performing x-ray diffraction experiments on the orthorhombic samples with incident x-rays both parallel and perpendicular to the stacking *b* direction, we clarified that the diffuse scattering peaks produced by these structures appear as streaks running toward the *b* axis direction, as shown in Figures 3*a*-3*c*. These diffuse streaks originate from the positions corresponding to *non*-integer indices; *e.g.,* (3/2 1 3/2), as shown in Figure 3*a*, indicating that short range ordering with a 2*a* × 2*c* superstructure emerges with weak correlation along the *b* axis. The *in*-plane correlation length was estimated to be almost 80 Å through the application of Scherrer's formula based on the measured cross-section width of the diffuse streaks. Based on these results, we can speculate that the short range ordering originates from the Fe/Pd arrangements in the *ac* plane with a correlation length of ~ 80 Å. One possible scenario is that Fe and Pd form checkerboard-like arrangements depending on the composition of $CaFe_{1-y}Pd_yAs_2$ with $y \cong 0.5$.

Temperature-independent diffuse scattering appeared in the tetragonal sample as well, probably as a result of the co-substitutions of Fe by Pd and Ca by La. Plate-like diffuse scattering appeared in the tetragonal samples, indicating a weak correlation. Further experimental studies will be required to clarify the accurate short range ordering patterns.

**XAFS experiments.** Although the tetragonal sample structure can possibly be expressed using the chemical formula $(Ca^{2+}_{1-x}La^{3+}_{x})(Fe^{2+}_{1-y}Pd^{2+}_{y}As^{3-})As^{-} \cdot xe^{-}$ with an excess charge of $xe^{-}$ injected into the $Fe^{2+}_{1-y}Pd^{2+}_{y}As^{3-}$ layers, we cannot easily predict the arsenic oxidation state for the orthorhombic sample owing to the short interlayer As-As length of $r_{As-As} \cong 2.47$ Å as shown in Figure 4*a*, which depicts the complex longitudinal network structures of arsenics and the resulting anomalous oxidation states of the system. To determine the bonding nature and oxidation states of the orthorhombic samples, we performed Fe and As K-edge XANES experiments for an orthorhombic sample with SrFe$_2$As$_2$, tetragonal and monoclinic structures, as described in the references. Figure 4*b* shows the Fe K-edge x-ray absorption near-edge spectra for the orthorhombic, tetragonal, and monoclinic samples. The overall appearances of these are quite similar, with a common prominent pre-edge peak at around 7110 eV that can be attributed to similar FeAs$_4$ tetragonal configurations. The edge energies are nearly identical, suggesting that quite similar oxidation states are present despite the different arsenic network structures. Figure 4*c* shows As K-edge XANES spectra for orthorhombic and SrFe$_2$As$_2$ with trivalent arsenics as a reference. The edge energies are consistent, clearly indicating the presence of trivalent arsenics consisting of Fe$_{1-y}$Pd$_y$As layers.

Table II. Crystallographic parameters of tetragonal and orthorhombic samples at 120 K based on refined site occupancy, atomic coordinates, and thermal parameters. $M(1)$ represents Ca$_{1-x}$La$_x$ with $x = 0.306(8)$. $M(2)$ and $M'(2)$ represent Fe$_{1-y}$Pd$_y$ with $y = 0.298(16)$ and $0.486(8)$, respectively. The occupancy is fixed at 1 in all atomic sites.

| | A. Tetragonal Sample | | |
|---|---|---|---|
| Refinement resolution (Å) | | | >0.40 |
| $R1$ (4σ cut) | | | 0.0583 |
| $R1$ (all reflections) | | | 0.0601 |
| Site | Atomic position | | |
| | x/a | y/b | z/c |
| As(1) | 0 | 0 | 0.32009(5) |
| As(2) | 0 | 0.5 | 0 |
| $M(1)$ | 0 | 0 | 0.11536(6) |
| $M(2)$ | 0 | 0.5 | 0.25 |
| | B. Orthorhombic sample | | |
| Refinement resolution (Å) | | | >0.40 |
| $R1$ (4σ cut) | | | 0.0295 |
| $R1$ (all reflections) | | | 0.0325 |
| Site | Atomic position | | |
| | x/a | y/b | z/c |
| As(1) | 0.5 | 0.03302(4) | 0.25 |
| As(2) | 0.5 | 0.17324(5) | 0.25 |
| Ca(1) | 0 | 0.10252(7) | 0.75 |
| $M'(2)$ | 0 | 0.25118(4) | 0.25 |

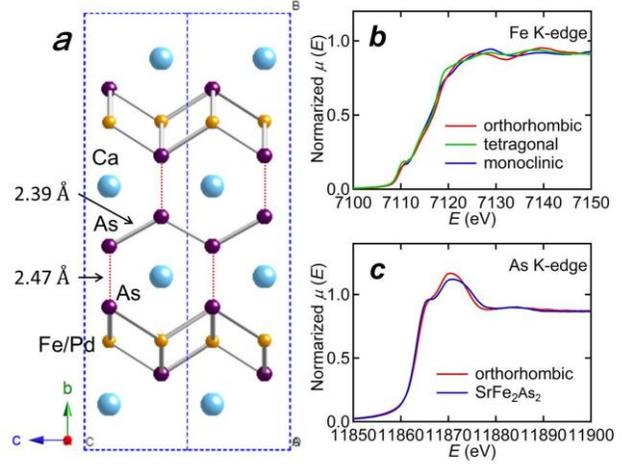

Figure 4. *a* Schematic of possible arsenic network structures in orthorhombic sample. *b* and *c* Absorption spectrum measured at Fe and As K-edge on orthorhombic sample with references. Tetragonal and monoclinic samples were used for *b*, and SrFe$_2$As$_2$ was employed for *c*, as references.

Based on our XANES experimental results, we can successfully deduce the chemical formula of the orthorhombic samples to be $Ca^{2+}(Fe^{2+}_{1-y}Pd^{2+}_{y}As^{3-})As^{-}$, with perhaps some excess charge carriers. This in turn implies the absence of strong longitudinal As-As bonding connecting the neighboring layers in the orthorhombic samples although the short interlayer expects us the presence of covalent bonding there. Some attractive interaction should work between the neighboring layers without changing the formal electron counts.

**Phase diagram and discussion.** As clarified from the synchrotron x-ray diffraction experiments, the samples consist of alternately stacked (Fe,Pd)As layers with arsenic network structures located between them. Bulk superconductivity has never been detected in tetragonal or orthorhombic samples, probably as a result of the oversubstitution of Pd for Fe. However, the existence of (Fe,Pd)As layers would suggest the emergence of superconductivity as a result of proper tuning of the substitution amounts of Pd. To determine the compositions of the examined 112-type arsenides, we performed additional single crystal x-ray diffraction experiments. Using single crystalline samples with different nominal compositions of Ca: La: Fe: Pd: As = 1-*x*: *x*: 1-*y*: *y*: 2, where $0 < x < 0.4$ and $0 < y < 0.6$, we clarified that both tetragonal and orthorhombic structures stabilize in a limited composition areas, as shown in Figure 5*a*. The monoclinic, tetragonal, and orthorhombic phases are obviously divided in each composition.

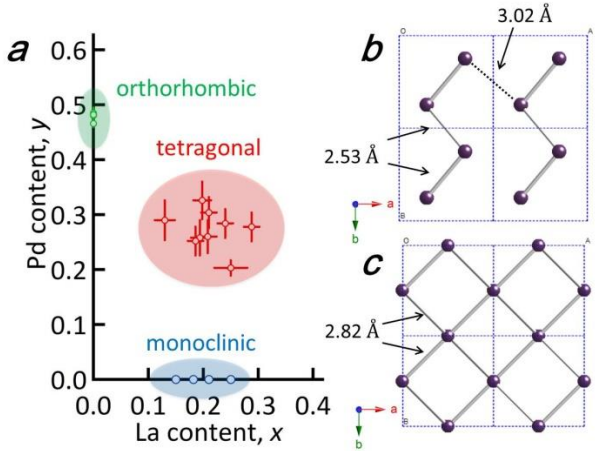

Figure 5. *a* shows the phase diagram as a function of the La content, *x*, and Pd content, *y*, obtained from the synchrotron x-ray diffraction experiments using several single crystals. *b* and *c* show the arsenic network structures for superconducting monoclinic and tetragonal samples, respectively.

Based on the differences in the lattice parameters among monoclinic, orthorhombic and tetragonal samples, we suggest that deformation of the arsenic network structures is mainly driven by lattice expansion caused by Pd substitution. As summarized in Table I, the *in*-plane lattice parameters increase under increased Pd content, *y*, mainly as a result of the ionic size effect in which a longer lattice parameter elongates and destabilizes the zigzag arsenic covalent bonding with $r_{As-As} \cong 2.53$ Å in the superconducting monoclinic phase, leading to the emergence of hypervalent networks with longer chemical bonding with $r_{As-As} \cong 2.82$ Å in the tetragonal phase, as shown in Figures 5*b* and 5*c*. Further elongation of the *in*-plane lattice parameters by additional Pd substitution makes the hypervalent networks unstable, resulting in abandonment of the *in*-plane two-dimensional network structures and the resulting formation of the longitudinal zigzag structures. We would like to note that the above relation can be also found in other 112-type materials. Whereas hypervalent square networks prefer *in*-plane lattice parameters that are longer than the zigzag chain layers in the 112-type $AnTSb_2$ (*An*: alkaline earth elements, *T*: 3*d* transition metals)[20-22], the contributions of *s-p* hybridization are stabilized in the longitudinal zigzag structures as a result of *in*-plane parameters that are larger than the corresponding parameters in the hypervalent square networks in 112-type silicides[23-26]. Quantitatively, the deformation of the arsenic network structures is more sensitive against lattice expansion effects in the present materials compared with other conventional 112-type materials, implying other factors help the deformation of arsenic network structures in the present arsenides, such as carrier doping effect by substituting Fe for Pd[27] and/or difference in electronegativity between Fe and Pd.

Finally, we would like to discuss possible 112-type high $T_c$ superconductors based on various arsenic network structures. As clarified by the XANES experiments, the formal oxidation states of the superconducting monoclinic, orthorhombic, and tetragonal samples were quite similar, suggesting potential superconductivity in 112-type iron arsenides with various arsenic network structures. Considering that 122-type $Sr(Fe_{1-x}Pd_x)_2As_2$ and $Ba(Fe_{1-x}Pd_x)_2As_2$ exhibit superconductivity within narrow *x* regions of $0.05 < x < 0.10$ and $0.021 < x < 0.077$ for $Sr(Fe_{1-x}Pd_x)_2As_2$[28] and $Ba(Fe_{1-x}Pd_x)_2As_2$[29], respectively, we can conclude that the oversubstitution of Pd for Fe should suppress the emergence of superconductivity in the examined samples. From this point of view, utilizing arsenic cis-trans chains may make it possible to produce novel 112-type high $T_c$ superconductivity. Cis-trans chains tend to prefer slightly longer lattice parameters than do the zigzag chain in $LnTAs_2$[30,31] (*Ln*: lanthanoide elements, *T*: 4*d* transition metals), suggesting that an as yet undescribed 112-type iron arsenide with cis-trans chains would require only a slight substitution of proper ions to produce superconductivity.

### ■ CONCLUSION

Taking advantage of the various network structures of arsenics, we successfully produced two previously unknown 112-type Pd doped iron arsenides. Their energy dependences on the anomalous dispersion terms of arsenic atomic form factors enabled us to determine accurate crystal structures for each. Considering that similar formal oxidation states were available among the monoclinic, orthorhombic, and tetragonal samples studied, proper tuning of substitution amounts allow further opportunities to investigate novel 112-type iron arsenide superconductors.

### ■ AUTHOR INFORMATION

#### Corresponding Author

* katayama@mcr.nuap.nagoya-u.ac.jp (N.K.).

**Notes**

The authors declare no competing financial interests.

■ **ACKNOWLEDGEMENT**

The authors acknowledge Dr. S. Kobayashi for the critical reading of the paper and Prof. M. Tabuchi for the technical support in XANES experiments. Part of this work was performed at the Department of Applied Physics, Nagoya University. It was partially supported by a Grant-in-Aid for JSPS KAKENHI Grant Number 15K17705 from the Japan Society for the Promotion of Science (JSPS), The Kurata Memorial Hitachi Science and Technology Foundation, and Nippon Sheet Glass Foundation for Materials Science and Engineering. The synchrotron radiation experiments at BL8A of PF, KEK were performed under the approval of the Photon Factory Program Advisory Committee (Proposal No. 2014S2-003). The XANES experiments were conducted at the BL11S2 of Aichi Synchrotron Radiation Center, Aichi Science & Technology Foundation, Aichi, Japan (Proposal No. 201605096).

■ REFERENCES

(1) Nagamatsu, J.; Nakagawa, N.; Muranaka, T.; Zenitani, Y.; Akimitsu, J. Superconductivity at 39 K in magnesium diboride. *Nature* **2001**, 410, 63.

(2) Emery, N.; Hérold, C.; d′Astuto, M.; Garcia, V.; Bellin, C.; Marêché, J.F.; Lagrange, P.; Loupias, G. Superconductivity of Bulk CaC6. *Phys. Rev. Lett.* **2005**, 95, 087003.

(3) Deller, V.K.; Eisenmann B. Darstellung und Kristallstruktur von CaSb2. *Z. anorg. allg. Chem.* **1976**, 425, 104-108.

(4) Hulliger, F.; Schmerczer R. Crystal structure and antiferromagnetism of EuSb2. *J. Solid State Chem.* **1978**, 26, 389-396.

(5) Papoian, G.A.; Hoffmann, R. Hypervalent Bonding in One, Two, and Three Dimensions: Extending the Zintl-Klemm Concept to Nonclassical Electron-Rich Networks. *Angew. Chem. Int. Ed.* **2000**, 39, 2408-2448.

(6) Kakiya, S.; Kudo, K.; Nishikubo, Y.; Oku, K.; Nishibori, E.; Sawa, H.; Yamamoto, T.; Nozaka, T.; Nohara, M.; Superconductivity at 38 K in Iron-Based Compound with Platinum-Arsenide Layers $Ca_{10}(Pt_4As_8)(Fe_{2-x}Pt_xAs_2)_5$. *J. Phys. Soc. Jpn.* **2011**, *80*, 093704.

(7) Nohara, M.; Kakiya, S.; Kudo, K.; Oshiro, Y.; Araki, S.; Kobayashi, T.C.; Oku, K.; Nishibori, E.; Sawa, H. Iron-platinum-arsenide superconductors $Ca_{10}(Pt_nAs_8)(Fe_{2-x}Pt_xAs_2)_5$. *Solid State Commun.* **2012**, *152*, 635-639.

(8) Ni, N.; Allred, J.M.; Chan, B.C.; Cava, R.J. High $T_c$ electron doped $Ca_{10}(Pt_3As_8)(Fe_2As_2)_5$ and $Ca_{10}(Pt_4As_8)(Fe_2As_2)_5$ superconductors with skutterudite intermediary layers. *Proc. Natl. Acad. Sci.* **2011**, *108*, E1019-E1026.

(9) Löhnert, C.; Stürzer, T.; Tegel, M.; Frankovsky, R.; Friederichs, G.; Johrendt, D. *Angew. Chem. Int. Ed.* **2011**, *50*, 9195-9199.

(10) Kudo, K.; Mitsuoka, D.; Takasuga, M.; Sugiyama, Y.; Sugawara, K.; Katayama, N.; Sawa, H.; Kubo, H.S.; Takamori, K.; Ichioka, M.; Fujii, T.; Mizokawa, T.; Nohara, M.; Superconductivity in $Ca_{10}(Ir_4As_8)(Fe_2As_2)_5$ with Square-Planar Coordination of Iridium. *Sci. Rep.* **2013**, *3*, 3101-1-5.

(11) Katayama, N.; Sugawara, K.; Sugiyama, Y.; Higuchi, T.; Kudo, K.; Mitsuoka, D.; Mizokawa, T.; Nohara, M.; Sawa, H. Synchrotron X-ray diffraction study of the Structural Phase Transition in $Ca_{10}(Ir_4As_8)(Fe_{2-x}Ir_xAs_2)_5$ . *J. Phys. Soc. Jpn.* **2014**, 8*3*, 113707.

(12) Hieke, C.; Lippmann, J.; Stürzer, T.; Friederichs, G.; Nitsche, F.; Winter, F.; Pöttgen, R.; Johrendt, D.; Superconductivity and crystal structure of the palladium–iron–arsenides Ca10(Fe1−xPdxAs)10Pd3As8. *Philos. Mag.* **2013**, 9*3*, 3680-3689.

(13) Katayama, N.; Kudo, K.; Onari, S.; Mizukami, T.; Sugawara, K.; Sugiyama, Y.; Kitahama, Y.; Iba, K.; Fujimura, K.; Nishimoto, N.; Nohara, M.; Sawa, H.; Superconductivity in $Ca_{1-x}La_xFeAs_2$: A Novel 112-Type Iron Pnictide with Arsenic Zigzag Bonds. *J. Phys. Soc. Jpn.* **2013**, *82*, 123702.

(14) Kudo, K.; Mizukami, T.; Kitahama, Y.; Mitsuoka, D.; Iba, K.; Fujimura, K.; Nishimoto, N.; Hiraoka, Y.; Nohara, M. Enhanced Superconductivity up to 43 K by P/Sb Doping of $Ca_{1-x}La_xFeAs_2$ *J. Phys. Soc. Jpn.* **2014**, *83*, 025001.

(15) Yakita, H.; Ogino, H.; Okada, T.; Yamamoto, A.; Kishio, K.; Tohei, T.; Ikuhara, Y.; Gotoh, Y.; Fujihisa, H.; Kataoka, K.; Eisaki, H.; Shimoyama, J. A New Layered Iron Arsenide Superconductor: (Ca,Pr)FeAs2 *J. Am. Chem. Soc.* **2014**, *136*, 846-849.

(16) Katayama, N.; Sugawara, K.; Nakano, A.; Kitou, S.; Sugiyama, Y. Kawaguchi, N. Ito, H.; Higuchi, T.; Fujii, T.; Sawa H. Synchrotron X-ray diffraction study of 112-type Ca1-xLaxFeAs2. *Physica C* **2015**, 518, 10-13.

(17) Rutzinger, D.; Bartsch, C.; Doerr, M.; Rosner, H.; Neu, V. Doert, Th.; Ruck, M. Lattice distortions in layered type arsenides *Ln*TAs2 (*Ln* = La-Nd, Sm, Gd, Tb; *T* = Ag, Au): Crystal structurres, electronic and magnetic properties. *J. Solid State Chem.* **2010**, 183, 510-520.

(18) Brechtel, E.; Cordier, G.; Schäfer, M. Neue ternäre erdalkali-übergangselement-pnictide. *J. Less-Common Met.* **1981**, 79, 131.

(19) Gribanov, A.; Grytsiv, A.; Rogl, P.; Seropegin, Y.; Giester, G. X-ray structural study of intermetallic alloys *RT*2Si and *RT*Si2 (*R* = rare earth, *T* = noble metal). *J. Solid State Chem.* **2010**, 183, 1278-1289.

(20) Brechtel, E.; Cordier, G.; Schäfer, H. Zur Darstellung und Kristallstruktur des SrZnSb2. *Z. Naturforsch.* **1979**, 34b, 251-255.

(21) Brechtel, E.; Cordier, G.; Schäfer, H. Neue ternäre erdalkali-übergangselement-pnictide. *J. Less-Common Met.*, **1978**, 79, 131-138.

(22) Farhan, M.A.; Lee, G.; Shim, J.H. AEMnSb2 (AE=Sr, Ba): a new class of Dirac materials. *J. Phys.: Condens. Matter*, **2014**, 26, 042201.

(23) Gribanov,A.; Grytsiv,A.; Rogl, P.; Seropegin, Y.; Giester, G. X-ray structural study of intermetallic alloys RT2Si and RTSi2 (R=rare earth, T=noble metal). *J. Solid State Chem.*, **2010**, 183, 829–843.

(24) Paccard, L.; Paccard, D.; Allemand, J. RFexSi2 (R ≡ Tb, Dy, Ho, Er, Lu) structures with the

non-stoichiometric TbFeSi2 type. *J. Less-Common Met.*, **1990**, 161, 295-298.
(25) Malaman, B.; Venturini, G.; Caër, G.L.; Pontonnier, L.; Fruchart, D.; Tomala, K.; Sanchez, J.P.; Magnetic structures of PrFeSi2 and NdFeSi2 from neutron and Mössbauer studies. *Phys. Rev. B*, **1990**, 41, 4700.
(26) Malaman, B.; Venturini, G.; Pontonnier, L.; Fruchart, D.; Magnetic structures of RMnSi2 compounds (R = La, Ce, Pr, Nd) from neutron study. *J. Magn. Magn. Mat.* **1986**, 53, 309.
(27) Ni, N.; Thaler, A.; Kracher, A.; Yan, J.Q.; Bud'ko, S.L., Canfield, P.C.; Phase diagrams of Ba($Fe_{1-x}M_x$)$_2$As$_2$ single crystals ($M$ = Rh and Pd). *Phys. Rev. B* **2009**, 80, 024511.
(28) Han, F.; Zhu, X.; Cheng, P.; Mu, G.; Jia, Y. Fang, L.; Wang, Y.; Luo, H.; Zeng, B.; Shen, B.; Shan, L.; Ren, C.; Wen, H-H.; Superconductivity and phase diagrams of the 4$d$ - and 5$d$ -metal-doped iron arsenides SrFe2−$xMx$As2 ($M$ = Rh, Ir, Pd). *Phys. Rev. B* **2009**, 80, 024506.
(29) Ni, N.; Thaler, A.; Kracher, A.; Yan, J.Q.; Bud′ko, S.L.; Canfield., P.C.; Phase diagrams of Ba(Fe1−$xMx$)2As2 single crystals ($M$=Rh and Pd) *Phys. Rev. B* **2009**, 80, 024511.
(30) Rutzinger, D.; Bartsch, C.; Doerr, M.; Rosner, H.; Neu, V.; Doert, Th.; Ruck, M.; Lattice distortions in layered type arsenides *LnT*As2 (*Ln* = La–Nd, Sm, Gd, Tb; *T* = Ag, Au): Crystal structures, electronic and magnetic properties. *J. Solid State Chem.*, **2010**, 183, 510-520.
(31) Demchyna, R.; Jemetio, J.P.F.; Prots, Y.; Doert, Th.; Akselrud, L.G.; Schnelle, W.; Kuz′ma, Y.; Grin, Y. CeAgAs2 - a New Derivative of the HfCuSi2 Type of Structure: Synthesis, Crystal Structure and Magnetic Properties. *Z. Anorg. Allg. Chem.* **2004**, 630, 635-641.